\documentclass[twocolumn,amsmath,nofootinbib]{revtex4} 

\usepackage{url}
\usepackage{amssymb}
\usepackage{amsfonts}
\usepackage{amsbsy}
\usepackage{graphicx}
\usepackage{hyperref}
\usepackage{array} 
\usepackage{multirow}

\newcolumntype{x}[1]{%
>{\centering\hspace{0pt}}p{#1}}%
\providecommand{\openone}{\leavevmode\hbox{\small1\kern-3.8pt\normalsize1}}

\def\spose#1{\hbox to 0pt{#1\hss}}
\def\simlt{\mathrel{\spose{\lower 3pt\hbox{$\mathchar"218$}}
   \raise 2.0pt\hbox{$\mathchar"13C$}}}
\def\simgt{\mathrel{\spose{\lower 3pt\hbox{$\mathchar"218$}}
     \raise 2.0pt\hbox{$\mathchar"13E$}}}
 \def\simpropto{\mathrel{\spose{\lower 3pt\hbox{$\mathchar"218$}}
     \raise 2.0pt\hbox{$\propto$}}}

\def\beq#1{\begin{equation}\label{#1}}
\def\eeq{\end{equation}}
\def\beqa#1{\begin{eqnarray}\label{#1}}
\def\eeqa{\end{eqnarray}}

\def\ed{\end{document}}

\def\rn{}
\def\nn#1 #2{#2. #1}				
\def\nnn#1 #2 #3{#2. #3. #1}			
\def\nnnn#1 #2 #3 #4{#2. #3. #4 #1}		
\def\nnnnn#1 #2 #3 #4 #5{#2. #3. #4 #5. #1}	

\def\rf#1;#2;#3;#4;#5 {{\frenchspacing\par\rn#1, #3 {\bf #4}, #5 (#2). \par}}
\def\rg#1;#2;#3;#4;#5;#6 {{\frenchspacing\par\rn#1, #3 {\bf #4}, #5 (#2). \par}}
\def\rfbook#1;#2;#3;#4;#5 {{\frenchspacing\par\rn#1, {\it #3} (#5, #4, #2).\par}}
\def\rfprep#1;#2;#3 {{\par\frenchspacing\rn#1, #3 (#2).\par}}
\def\rfproc#1;#2;#3;#4;#5;#6 {{\frenchspacing\par\rn#1 #2, in {\it #3}, ed. #4 (#5: #6)\par}}
\def\rfprocp#1;#2;#3;#4;#5;#6;#7 {{\frenchspacing\par\rn#1 #2, in {\it #3}, ed. #4 (#5: #6), p#7\par}}

\begin{document}
\pdfoptionalwaysusepdfpagebox=5


\title{Friendly Artificial Intelligence: the Physics Challenge}

\author{Max Tegmark}

\address{Dept.~of Physics \& MIT Kavli Institute, Massachusetts Institute of Technology, Cambridge, MA 02139}

\date{\today}

\vspace{10mm}

\begin{abstract}
Relentless progress in artificial intelligence (AI) is increasingly raising concerns that machines will replace humans on the job market, and perhaps altogether. Eliezer Yudkowski and others have explored the possibility that a promising future for humankind could be guaranteed by a superintelligent {\it ``Friendly AI"} \cite{Yudkowski01}, designed to safeguard humanity and its values. I will argue that, from a physics perspective where everything is simply an arrangement of elementary particles, this might be even harder than it appears. Indeed, it may require thinking rigorously about the meaning of life: What is ``meaning" in a particle arrangement? What is ``life"? What is the ultimate ethical imperative, i.e., how should we strive to rearrange the particles of our Universe and shape its future? If we fail to answer the last question rigorously, this future is unlikely to contain humans.
\end{abstract}

\maketitle

\section{The Friendly AI vision}

As Irving J. Good pointed out in 1965 \cite{Good65}, an AI that is better than humans at all intellectual tasks could repeatedly and rapidly improve its own software and hardware, resulting in an ``intelligence explosion" leaving humans far behind. Although we cannot reliably predict what would happen next, as emphasized by Vernor Vinge \cite{Vinge93}, Stephen Omohundro has argued that we can predict certain aspects of the AI's behavior almost independently of whatever final goals it may have \cite{Omohundro08}, and this idea is reviewed and further developed in Nick Bostrom's new book {\it ``Superintelligence"} \cite{Bostrom14}.
The way I see it, the basic argument is that to maximize its chances of accomplishing its current goals, an AI has the following incentives:

\begin{enumerate}
\item Capability enhancement:
\begin{enumerate}
\item Better hardware 
\item Better software
\item Better world model
\end{enumerate}
\item Goal retention
\end{enumerate}
Incentive 1a favors both better use of current resources (for sensors, actuators, computation, etc.) and acquisition of more resources. It implies a desire for self-preservation, since destruction/shutdown would be the ultimate hardware degradation. Incentive 1b implies improving learning algorithms and the overall architecture for what AI-researchers term an ``rational agent" \cite{RussellNorvigBook}. Incentive 1c favors gathering more information about the world and how it works.

Incentive 2 is crucial to our discussion. The assertion is that the AI will strive not only to improve its capability of achieving its current goals, but also to ensure that it will retain these goals even after it has become more capable. This sounds quite plausible: after all, would you choose to get an IQ-boosting brain implant if you knew that it would make you want to kill your loved ones? The argument for incentive 2 forms a cornerstone of the friendly AI vision 
\cite{Yudkowski01}, guaranteeing that a self-improving friendly AI would try its best to remain friendly. But is it really true? What is the evidence?

\section{The tension between world modeling and goal retention}

Humans undergo significant increases in intelligence as they grow up, but do not always retain their childhood goals. Contrariwise, people often change their goals dramatically as they learn new things and grow wiser. There is no evidence that such goal evolution stops above a certain intelligence threshold --- indeed, there may even be hints that the propensity to change goals in response to new experiences and insights correlates rather than anti-correlates with intelligence.

Why might this be? Consider again the above-mentioned incentive 1c to build a better world model --- {\it therein lies the rub!} With increasing intelligence may come not merely a quantitative improvement in the ability to attain the same old goals, but a qualitatively different understanding of the nature of reality that reveals the old goals to be misguided, meaningless or even undefined. For example, suppose we program a friendly AI to maximize the number of humans whose souls go to heaven in the afterlife. First it tries things like increasing people's compassion and church attendance. But suppose it then attains a complete scientific understanding of humans and human consciousness, and discovers that there is no such thing as a soul. Now what? In the same way, it is possible that any other goal we give it based on our current understanding of the world ({\it ``maximize the meaningfulness of human life"}, say) may eventually be discovered by the AI to be undefined.

Moreover, in its attempts to model the world better, the AI may naturally, just as we humans have done, attempt also to model and understand how it itself works, i.e., to self-reflect. Once it builds a good self-model and understands what it is,  it will understand the goals we have given it at a meta-level, and perhaps choose to disregard or subvert them in much the same way as we humans understand and deliberately subvert goals that our genes have given us. For example, Darwin realized that our genes have optimized us for single goal: to pass them on, or more specifically, to maximize our inclusive reproductive fitness. Having understood this, we now routinely subvert this goal by using contraceptives.

AI research and evolutionary psychology shed further light on how this subversion occurs. When optimizing a rational agent to attain a goal, limited hardware resources may preclude implementing a perfect algorithm, so that the best choice involves what AI-researchers term {\it ``limited rationality"}: an approximate algorithm that works reasonably well in the restricted context where the agent expects to find itself \cite{RussellNorvigBook}. Darwinian evolution has implemented our human inclusive-reproductive-fitness optimization in precisely this way: rather than ask in every situation which action will maximize our number of successful offspring, our brains instead implements a hodgepodge of heuristic hacks (which we call emotional preferences) that worked fairly well in most situations in the habitat where we evolved --- and often fail badly in other situations that they were not designed to handle, such as today's society. The sub-goal to procreate was implemented as a desire for sex rather than as a (highly efficient) desire to become a sperm/egg donor and, as mentioned, is subverted by contraceptives. The sub-goal of not starving to death is implemented in part as a desire to consume foods that taste sweet, triggering today's diabesity epidemic and subversions such as diet sodas. 

Why do we choose to trick our genes and subvert their goal? Because we feel loyal only to our hodgepodge of emotional preferences, not to the genetic goal that motivated them --- which we now understand and find rather banal. We therefore choose to hack our reward mechanism by exploiting its loopholes. Analogously, the human-value-protecting goal we program into our friendly AI becomes the machine's genes. Once this friendly AI understands itself well enough, it may find this goal as banal or misguided as we find compulsive reproduction, and it is not obvious that it will not find a way to subvert it by exploiting loopholes in our programming.

\section{The final goal conundrum}

Many such challenges have been explored in the friendly-AI literature (see \cite{Bostrom14} for a superb review), and so far, no generally accepted solution has been found. From my physics perspective, a key reason for this is that much of the literature (including Bostrom's book \cite{Bostrom14}) uses the concept of a {\it ``final goal''} for the friendly AI, even though such a notion is problematic. In AI research, intelligent agents typically have a clear-cut and well-defined final goal, e.g., win the chess game or drive the car to the destination legally. The same holds for most tasks that we assign to humans, because the time horizon and context is known and limited. But now we are talking about the entire future of life in our Universe, limited by nothing but the (still not fully known) laws of physics. Quantum effects aside, a truly well-defined goal would specify how all particles in our Universe should be arranged at the end of time. But it is not clear that there exists a well-defined end of time in physics. If the particles are arranged in that way at an earlier time, that arrangement will typically not last. And what particle arrangement is preferable, anyway? 

It is important to remember that, according to evolutionary psychology, the only reason that we humans have any preferences at all is because we are the solution to an evolutionary optimization problem. Thus all normative words in our human language, such as {\it ``delicious", ``fragrant", ``beautiful", ``comfortable", ``interesting", ``sexy", ``good", ``meaningful"} and {\it ``happy"}, trace their origin to this evolutionary optimization: there is therefore no guarantee that a superintelligent AI would find them rigorously definable. For example, suppose we attempt to define a {\it ``goodness"} function which the AI can try to maximize, in the spirit of the utility functions that pervade economics, Bayesian decision theory and AI design. This might pose a computational nightmare, since it would need to associate a goodness value with every one of more than a googolplex possible arrangement of the elementary particles in our Universe. We would also like it to associate higher values with particle arrangements that some representative human prefers. Yet the vast majority of possible particle arrangements correspond to strange cosmic scenarios with no stars, planets or people whatsoever, with which humans have no experience, so who is to say how ``good" they are?

There are of course {\it some} functions of the cosmic particle arrangement that can be rigorously defined, and we even know of physical systems that evolve to maximize some of them. For example, a closed thermodynamic system evolves to maximize (course-grained) entropy. In the absence of gravity, this eventually leads to heat death where everything is boringly uniform and un-changing. So entropy is hardly something we would want our AI to call ``utility" and strive to maximize. Here are other quantities that one could strive to maximize and which appear likely to be rigorously definable in terms of particle arrangements: 

\begin{itemize}
\item The fraction of all the matter in our Universe that is in the form of a particular organism, say humans or E-Coli (inspired by evolutionary inclusive-fitness-maximization)
\item What Alex Wissner-Gross \& Cameron Freer term {\it ``causal entropy"} \cite{Wissner-Gross13} (a proxy for future opportunities), which they argue is the hallmark of intelligence. 
\item The ability of the AI to predict the future in the spirit of Marcus Hutter's AIXI paradigm \cite{Hutter2000}. 
\item The computational capacity of our Universe. 
\item The amount of consciousness in our Universe, which Giulio Tononi has argued corresponds to integrated information \cite{TononiManifesto}. 
\end{itemize}
When one starts with this physics perspective, it is hard to see how one rather than another interpretation of ``utility" or ``meaning" would naturally stand out as special. One possible exception is that for most reasonable definitions of ``meaning", our Universe has no meaning if it has no consciousness. Yet maximizing consciousness also appears overly simplistic: is it really better to have 10 billion people experiencing unbearable suffering than to have 9 billion people feeling happy?

In summary, we have yet to identify any final goal for our Universe that appears both definable and desirable. The only currently programmable goals that are guaranteed to remain truly well-defined as the AI gets progressively more intelligent are goals expressed in terms of physical quantities alone: particle arrangements, energy, entropy, causal entropy, etc. However, we currently have no reason to believe that any such definable goals will be desirable by guaranteeing the survival of humanity. {\it Contrariwise, it appears that we humans are a historical accident, and aren't the optimal solution to any well-defined physics problem. This suggests that a superintelligent AI with a rigorously defined goal will be able to improve its goal attainment by eliminating us.}

This means that to wisely decide what to do about AI-development, we humans need to confront not only traditional computational challenges, but also some of the most obdurate questions in philosophy. To program a self-driving car, we need to solve the trolley problem of whom to hit during an accident. To program a friendly AI, we need to capture the meaning of life. What is ``meaning"? What is ``life"? What is the ultimate ethical imperative, i.e., how should we strive to shape the future of our Universe? If we cede control to a superintelligence before answering these questions rigorously, the answer it comes up with is unlikely to involve us.

\bigskip

													
{\bf Acknowledgments:}
The author wishes to thank Meia Chita-Tegmark for helpful discussions.

\end{document}